\documentclass[final]{siamltex}
\usepackage{amssymb}
\usepackage{amsmath}
\usepackage{epsfig}
\usepackage{psfrag}
\usepackage{subfigure}

\renewcommand{\d}{\mathrm{d}}
\newcommand{\Hin}{H^{\mathrm{in}}}
\newcommand{\Hout}{H^{\mathrm{out}}}
\newcommand{\R}{\mathbb{R}}
\newcommand{\rar}{\rightarrow}
\newtheorem{defn}{Definition}
\newcommand{\Fix}{\mathrm{Fix}\ }
\newcommand{\Poincare}{Poincar\'{e} }
\newcommand{\e}{\mathrm{e}}
\newcommand{\etal}{\textit{et al.\ }}
\newcommand{\ie}{i.e.\ }
\newcommand{\Z}{\mathbb{Z}}
\renewcommand{\vec}[1]{\mathbf{#1}}

\graphicspath{{../../../figures/}{figures/}}

\begin{document}

\title{Stabilisation of long-period periodic orbits using time-delayed feedback control}
\author{Claire M. Postlethwaite\thanks{Department of Mathematics, University of Auckland, Private Bag 92019, Auckland, New Zealand, and Engineering Science and Applied Mathematics, Northwestern University, Evanston, IL, 60208, USA, {\tt c.postlethwaite@math.auckland.ac.nz}.}}
\maketitle

\begin{abstract}
The Pyragas method of feedback control has attracted much interest as a method of stabilising unstable periodic orbits in a number of situations. We show that a time-delayed feedback control similar to the Pyragas method can be used to stabilise 
periodic orbits with arbitrarily large period, specifically those resulting from a 
resonant bifurcation of a heteroclinic cycle. Our analysis reduces the infinite-dimensional delay-equation governing the system with feedback to a three-dimensional map, by making certain assumptions about the form of the solutions. 
The stability of a fixed point in this map corresponds to the stability of the periodic orbit in the flow, and can be computed analytically. We compare the analytic results to a numerical example and find very good agreement.
\end{abstract}



\pagestyle{myheadings}
\thispagestyle{plain}
\markboth{CLAIRE M. POSTLETHWAITE}{STABILISATION OF LONG-PERIOD PERIODIC ORBITS}

\section{Introduction}

The stabilisation of unstable periodic orbits (UPOs) using feedback
control has attracted the attention of many authors over a number of years.
The time-delayed feedback method of Pyragas~\cite{Pyr92}, has
been of particular interest. Here, the 
feedback $F$ is proportional to the difference between the current and
a past state of the system. Specifically, $F=K(x(t-\tau)-x(t))$ where
$x(t)$ is some state vector, $\tau$ is the period of the
targeted UPO and $K$ is a feedback gain matrix. Advantages of this
method include the following. First, since the feedback vanishes on
any orbit with period $\tau$,
the targeted UPO is still a solution of the system
with feedback. Control is therefore achieved in a non-invasive manner.
Second, the only information required a priori is the period $\tau$
of the target UPO, rather than a detailed knowledge of the profile of
the orbit, or even any knowledge of the form of the original ODEs,
which may be useful in experimental setups. The method has been implemented successfully
in a variety of laboratory situations~\cite{PT93,GSCS94,BDG94,PBA96,FSK02,SBFHLM93,LFS95}, 
as well as analytically and numerically in spatially extended pattern-forming systems~\cite{BS96b, MS04,
  LYH96, PS06}; more examples can be found in a
recent review by Pyragas~\cite{Pyr06}.

Until now, there
has been little or no study on whether there are limitations to Pyragas feedback control as the period of the targeted orbit, and hence the delay time, becomes large. In this paper, we investigate the use of Pyragas feedback on unstable periodic orbits with arbitrarily large period.

One mechanism for the generation of long-period periodic orbits is at bifurcations from homoclinic orbits or heteroclinic cycles. In this paper we focus on a subcritical bifurcation from a 
symmetric heteroclinic cycle, specifically the heteroclinic cycle of Guckenheimer and Holmes~\cite{GH88}. The bifurcation produces a branch of unstable long-period periodic
orbits and we investigate using a time-delayed feedback control similar to the Pyragas feedback as a stabilisation mechanism. 

The addition of Pyragas feedback to the ODEs considered by Guckenheimer and Holmes results in an infinite-dimensional delay equation. In order to analyse trajectories near the periodic orbit of interest, we make a number of assumptions about the form of  solutions to the delay-differential equation and reduce the flow to a three-dimensional map. This method, after the assumptions have been made, is a modified version of the standard `small box and Poincar\'e map' analysis used by many authors to study the dynamics of trajectories close to heteroclinic cycles. This reduction of an infinite-dimensional delay equation to a finite dimensional map has not appeared before in the literature. Although our assumptions are not fully rigorously justified, we test the validity of our arguments by comparing our results with a numerical example. We find excellent agreement between the analytical and numerical results.
A surprising result of the analysis for the particular example we use is that as the period of the orbit increases, the amplitude of the gain parameter required to stabilise the unstable orbits decreases.

This paper is organised as follows. In section~\ref{sec:rev} we give a review of heteroclinic cycles and their bifurcations. We describe the Guckenheimer--Holmes heteroclinic cycle, and summarise the standard approach to analysing trajectories close to heteroclinic cycles. In section~\ref{sec:feed} we describe how we choose the feedback control terms which are added to the equations. We then perform the reduction of the equations described above, which gives us a method of computing the stability of the periodic orbits. Section~\ref{sec:num} contains numerical examples and section~\ref{sec:conc}
 concludes.

\section{Review of heteroclinic cycles}
\label{sec:rev}

A heteroclinic cycle is a topological circle of connecting orbits
between at least two saddle-type equilibria. In generic (non-symmetric) dynamical
 systems, heteroclinic cycles are of high codimension and their
existence for open sets of parameter values is unexpected.
If a dynamical system contains flow-invariant subspaces, the
connecting orbits can be contained within these subspaces, and then the
heteroclinic cycle is robust to perturbations of the system that
preserve the invariance of these subspaces. Flow invariant subspaces
can arise due to symmetry,  or due to other restrictions on the flow
(such as extinctions in population 
 dynamics models~\cite{HS98}). The review of Krupa~\cite{Kru97}
 contains many examples of robust heteroclinic cycles. 
In this paper we consider robust heteroclinic cycles in
 symmetric systems.

\subsection{Preliminary definitions}

Consider a continuous-time dynamical system defined by an ODE:
\begin{equation} \label{eq:ode1}\dot{x}=f(x),\qquad
x\in\R^n,\end{equation} 
where $f:\R^n\rar\R^n$ is a $\Lambda$-equivariant vector field, that is,
\begin{equation}\label{eq:equiv1}
\gamma f(x)=f(\gamma x),\qquad\forall\ \gamma\in\Lambda,\end{equation}
and $\Lambda\subset \vec{O}(n)$ is a finite Lie group.
An equilibrium $\xi\in\R^n$ of~\eqref{eq:ode1} satisfies $f(\xi)=0$. We
consider only hyperbolic equilibria, and assume that $f$ is smoothly
linearisable about each equilibrium.

\begin{defn}\label{def:het_con}
$\phi_j(t)$ is a \emph{heteroclinic connection} between two equilibria
$\xi_j$ and $\xi_{j+1}$ of~\eqref{eq:ode1} if $\phi_j(t)$
is a solution of~\eqref{eq:ode1} which is backward asymptotic to
$\xi_j$ and forward asymptotic to $\xi_{j+1}$.
\end{defn}

A heteroclinic cycle is an invariant set $X\subset\R^n$ consisting of the
union of a set of equilibria $\{\xi_1,...,\xi_m\}$ and
orbits $\{\phi_1,...,\phi_m\}$, where $\phi_j$ is a
heteroclinic connection between $\xi_j$ and $\xi_{j+1}$; and
$\xi_{m+1}\equiv\xi_1$. We require that $m\geq 2$. If $m=1$, then
$\phi_1$ is a homoclinic orbit. 
A heteroclinic cycle is a \emph{homoclinic cycle} if there exists
$\gamma\in\Lambda$ such that 
$\gamma\xi_j=\xi_{j+1}$ for all $j$.

For $x\in\R^n$ we define the \emph{isotropy subgroup} $\Sigma_x$,
\begin{equation}\label{eq:iso_sg} \Sigma_x=\{\sigma\in\Lambda:\sigma
  x=x\}.\end{equation} 
For $\Sigma$ an isotropy subgroup of $\Lambda$, we define the
\emph{fixed-point subspace}
\begin{equation} \label{eq:fps} \Fix\Sigma=\{x\in\R^n:\sigma x=x\
  \forall \sigma\in\Sigma\}.\end{equation} 

\begin{defn}\label{def:het_cyc}
A heteroclinic cycle $X$ is \emph{robust} if for each $j$, $1\leq
j\leq m$, there exists
a fixed-point subspace, $P_j=\Fix\Sigma_j$ where
$\Sigma_j\subset\Lambda$ and
\begin{enumerate}
\item $\xi_{j}$ is a saddle and $\xi_{j+1}$ is a sink for the flow
restricted to $P_j$,
\item there is a heteroclinic connection from $\xi_{j}$ to $\xi_{j+1}$
  contained in $P_j$.
\end{enumerate}
\end{defn}

Robust heteroclinic cycles occur as codimension-zero phenomena in
systems with symmetry. That is,  they can exist for open sets of parameter values.
 Bifurcations of heteroclinic cycles therefore occur as codimension-one phenomena.
 We now consider the computation of the stability of heteroclinic cycles and the 
associated bifurcations.

\subsection{Resonant bifurcations}

The stability of a heteroclinic cycle is usually computed by constructing \Poincare maps
 on a \Poincare section of the flow. The flow near the cycle is divided into two parts; the `local' 
part, near the equilibria, where the flow can be well approximated by the linearised flow 
about the equilibria, and the `global' part of the flow, where the trajectory is away 
from the equilibria. The global part of the flow occurs on a much faster timescale 
than the local part and can be approximated by a linearisation of the flow around 
the heteroclinic connections. The construction of such \Poincare maps is a standard
 procedure, details can be found in, for example~\cite{KS94, KM04}.

Heteroclinic cycles generically lose stability in two ways: resonant bifurcations and 
transverse bifurcations. Transverse bifurcations occur when one of the 
eigenvalues at an equilibrium passes through zero; the equilibrium 
undergoes a local
bifurcation. We do not consider transverse bifurcations here, 
see~\cite{Chos97} for details. Throughout this paper, when we refer to `the eigenvalues at an equilibrium', we of course mean the eigenvalues of the Jacobian matrix of the flow linearised about that equilibrium.

At a resonant bifurcation the eigenvalues at the equilibria are generically non-zero, but satisfy an algebraic condition that determines a
global change in the stability properties of the cycle. Resonant bifurcations were
first studied in the non-symmetric case by Chow et~al.~\cite{CDF90} in the context of a
bifurcation from a homoclinic orbit. A more recent study~\cite{PD06} considers a 
codimension-two resonant bifurcation from a robust heteroclinic cycle with complex eigenvalues.

Resonant bifurcations are generically accompanied by the birth or death of a long-period
 periodic orbit. If $\nu$ is the bifurcation parameter controlling the resonant bifurcation (that is,
 $\nu=0$ at the bifurcation point), then the period $T$ of the bifurcating periodic orbit generically scales as
\[T\sim\frac{1}{\nu}.\]
Resonant bifurcations can occur in a supercritical or subcritical manner. 
We consider the subcritical case, when the branching periodic orbits are unstable,
 and in the following show that Pyragas-type time-delayed feedback can stabilise the periodic orbits.
Our analysis focuses on the Guckenheimer--Holmes cycle in $\R^3$.

\subsection{The Guckenheimer--Holmes cycle}

The Guckenheimer--Holmes cycle~\cite{GH88} is a prototypical example of a robust heteroclinic 
cycle. We use this cycle as an example on which to base our analysis. First we review the 
original case with no feedback.

\begin{figure}
\psfrag{x1}{$\xi_1$}
\psfrag{x2}{$\xi_2$}
\psfrag{x3}{$\xi_3$}
\begin{center}
\epsfig{figure=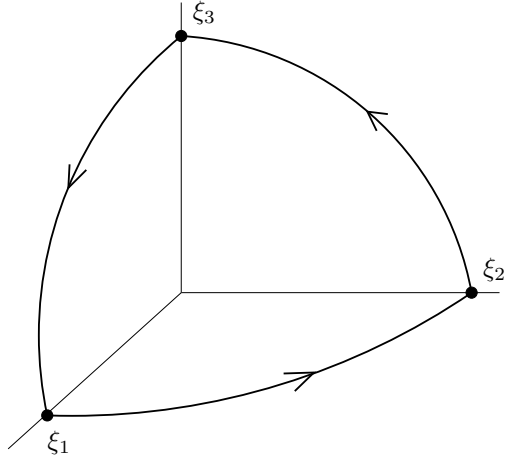,height=6cm}
\caption{The Guckenheimer--Holmes cycle in $\R^3$.}
\label{fig:GHcyc}
\end{center}
\end{figure}

The equations considered by Guckenheimer and Holmes can be written:
\begin{equation}
\begin{split}\label{eq:GH}
\dot{x}_1 &=x_1(1-\vec{X}-\mu x_2^2+\lambda x_3^2), \\
\dot{x}_2 &=x_2(1-\vec{X}-\mu x_3^2+\lambda x_1^2), \\
\dot{x}_3 &=x_3(1-\vec{X}-\mu x_1^2+\lambda x_2^2), 
\end{split}\end{equation}
where $\vec{X}=\sum_{i=1}^3 x_i^2$, and $\mu$ and $\lambda$ are real parameters. 
The equations are equivariant under the symmetry group 
$\Lambda=\Z_3\ltimes\Z_2^3$, generated by a reflection $\kappa_1$ and
a rotation $\gamma$:
\begin{align*}
\kappa(x_1,x_2,x_3)&=(-x_1,x_2,x_3),\\
\gamma(x_1,x_2,x_3)&=(x_3,x_1,x_2).
\end{align*} 
We label the equilibrium on the positive $x_j$-axis as $\xi_j$. Here, and throughout the
remainder of the paper, subscripts on equilibria, coordinates and similar objects should be taken mod $3$.
Each two-dimensional coordinate plane is a fixed point subspace. If $\mu\lambda>0$, then the only equilibria in each coordinate plane are those lying on the coordinate axes. We consider the case $\mu,\lambda>0$ and then it can be shown that in the plane $x_3=0$, $\xi_1$ is a saddle and $\xi_2$ is a sink. It can additionally be shown that in forward time trajectories are bounded away from infinity and therefore by the \Poincare--Bendixson theorem there exists a heteroclinic connection from $\xi_1$ to $\xi_2$. Similarly, connections also exist from $\xi_2$ to $\xi_3$, and $\xi_3$ to $\xi_1$.
These connections lie in two-dimensional fixed-point subspaces (the two-dimensional coordinate planes), so the cycle is robust. 
The resulting heteroclinic cycle
is shown schematically in figure~\ref{fig:GHcyc}. Also note that $\gamma\xi_j=\xi_{j+1}$, 
so the cycle is homoclinic.

The stability of the cycle can be calculated using the methods described above. It is a 
standard procedure, but we outline the method here, as we use similar ideas later when considering the stability of periodic orbits in the system with added time-delayed 
feedback. Consider a 
trajectory which passes close to the equilibrium $\xi_1$. The linearised flow near $\xi_1$ is:
\begin{align}
\dot x_1&=-2 x_1, \\
\dot x_2&=\lambda x_2, \\
\dot x_3&=-\mu x_3.
\end{align}
The $x_1$ direction is the `radial' direction, and as shown in~\cite{KM95}, for heteroclinic cycles of this type, the radial direction does not affect the stability of the cycle. All trajectories move away from the origin, and also away from infinity, and in this case are attracted to an `invariant sphere'~\cite{Field96} which contains the heteroclinic cycle. Therefore, for simplicity, we henceforth ignore this component.
We define \Poincare sections
close to $\xi_1$:
\begin{align}
\Hin_1&=\{(x_1,x_2,x_3)|\ 0<x_2<h, x_3=h\}, \\
\Hout_1&=\{(x_1,x_2,x_3)|\ x_2=h, 0<x_3<h \},
\end{align}
where $0<h\ll 1$, and construct a \Poincare return map on $\Hin_1$. Consider a trajectory which passes through $H_1^{\mathrm{in}}$ at time
 $t=0$ with $x_2(0)=x_2^i$. The trajectory will hit $H_1^{\mathrm{out}}$ at 
\[t=T_i\approx (-1/\lambda)\log(x_2^i/h)\]
with 
\[x_3(T_i)\equiv x_3^o\approx h^{1+\mu/\lambda}{x_2^i}^{\mu/\lambda}.\]
We thus write down a local map $\phi_{\mathrm{loc}}: H_1^{\mathrm{in}} \mapsto
H_1^{\mathrm{out}}$, which describes the flow near the equilibria:
\begin{equation}\label{eq:loc}
x_3^o=\phi_{\mathrm{loc}}(x_2^i)=h^{1+\mu/\lambda}{x_2^i}^{\mu/\lambda}.\end{equation}
The flow near the heteroclinic connection 
from $H_1^{\mathrm{out}}$ to a plane $\Hin_2=\gamma\Hin_1$ near
$\xi_2$ is approximated by the global map $\phi_{\mathrm{far}}$: 
\begin{equation} \label{eq:far}
x_3^i=\phi_{\mathrm{far}}(x_3^o)= Ax_3^o+O({x_3^o}^2)\end{equation}
where $x_3^i$ is the $x_3$ coordinate of the trajectory when it next hits
$H_2^{\mathrm{in}}$, and $A$ is a positive constant. Note that the constant term in this expansion of $\phi_{\mathrm{far}}$ is zero because the plane $x_3=0$ is invariant. 
We write $\tilde{\phi}=\phi_{\mathrm{far}}\circ\phi_{\mathrm{loc}}$. Since $\Hin_1=\gamma^{-1}\Hin_2$, the map $\gamma^{-1}\tilde{\phi}(x_2^i)$ is a return map on $\Hin_1$.
Write $\phi=\gamma^{-1}\tilde{\phi}$ and then the return map is
\[\phi(x)=Ax^{\delta}\]
where $\delta=\mu/\lambda$.

The map $\phi(x)$ has fixed points at $x=0$ and at
$x=x_p=A^{1/(1-\delta)}$. The fixed point at $x=0$ corresponds to
the heteroclinic cycle in the flow and is stable if $\delta>1$. The heteroclinic cycle loses stability in a resonant 
bifurcation at $\delta=1$. The second fixed point at $x=x_p$ corresponds to a branch of periodic
orbits, as long as $x_p$ is small and positive. 
The symmetry $\gamma$ acts as a spatio-temporal symmetry on the periodic orbits. That is, if we write the periodic solution as a trajectory $x^{\star}(t;\mu,\lambda)$, with minimal period $T$, then 
\[
 \gamma^{-1}x^{\star}(t)=x^{\star}(t-T/3).
\]

The stability of the orbits can be determined by finding the stability of the fixed point $x_p$ in
the map $\phi$. It is simple to see that if $A>1$, $x_p$ is small and positive (and
hence corresponds to a periodic orbit in the flow) when
$\delta>1$, so the resonant bifurcation is subcritical. We find that 
\[\frac{\d \phi}{\d x}_{x=x_p}=\delta,\]
and so $x_p$ is unstable. Conversely, if $A<1$, then $x_p$ corresponds to a branch of stable periodic orbits if $\delta<1$, and the
bifurcation is supercritical. The period $T$ of the orbit is approximately 
\[T=3\tau\approx \frac{-3\log A}{\lambda-\mu},\]
where $\tau=-\log(x_p)/\lambda$ and is the time spent by the trajectory each time it passes close to an equilibrium. We are ignoring the time spent away from the equilibria (that is, close to the heteroclinic
 connections in the invariant planes) since it is much less than $T$ when we
 are close to the resonant bifurcation, that is, $|1-\delta|\ll 1$.

In equations~\eqref{eq:GH}, the resonant heteroclinic bifurcation at $\mu=\lambda$ is degenerate. That is, the branch of periodic orbits exists only at $\mu=\lambda$. This corresponds to the case $A=1$ in the map $\phi$. We add additional higher order 
terms to break this degeneracy, specifically we consider
\begin{equation}
\begin{split}\label{eq:GH_alpha}
\dot{x}_1 &=x_1(1-\vec{X}-\mu x_2^2+\lambda x_3^2+\alpha x_2^2x_3^2), \\
\dot{x}_2 &=x_2(1-\vec{X}-\mu x_3^2+\lambda x_1^2+\alpha x_3^2x_1^2), \\
\dot{x}_3 &=x_3(1-\vec{X}-\mu x_1^2+\lambda x_2^2+\alpha x_1^2x_2^2). 
\end{split}\end{equation}
The additional terms preserve the equilibria and the symmetries of the
system, and also the invariant planes and the heteroclinic cycle. The heteroclinic
cycle still loses stability in a resonant bifurcation at
$\mu=\lambda$, but now a branch of periodic orbits is created in either $\mu>\lambda$ or $\mu<\lambda$ . The sign of $\alpha$ determines the branching direction and whether, in the map $\phi$, $A$ is greater or less than $1$.
 If $\alpha>0$, we see a branch of unstable periodic orbits in $\mu>\lambda$ (and the resonant bifurcation is subcritical).
 If $\alpha<0$, we see a branch of stable 
periodic orbits in $\mu<\lambda$ (and the bifurcation is supercritical). A complete study of the effect of fifth order terms on the dynamics near the GH cycle has not been performed. However, the above assertion can be seen by considering the effect of the new term on the $x_j$ component when the trajectory is close to the $x_j=0$ plane but away from either coordinate axis.

In the following, we consider the subcritical case, where the periodic orbits are
 unstable, and add non-invasive time-delayed feedback  to stabilise the orbits 
near the heteroclinic cycle.

\section{Addition of feedback terms}
\label{sec:feed}

\subsection{Choice of coordinates}

To ease analysis and improve the accuracy in the numerical computations in section~\ref{sec:num}, we introduce new coordinates
$Y_j=\log(x_j^2)$. Along with a change in timescale, this transforms equations~\eqref{eq:GH_alpha} to 
\begin{equation}
\begin{split}\label{eq:GH_log}
\dot{Y}_1 &=(1-\vec{Y}-\mu \e^{Y_2}+\lambda \e^{Y_3}+\alpha \e^{Y_2+Y_3}), \\
\dot{Y}_2 &=(1-\vec{Y}-\mu \e^{Y_3}+\lambda \e^{Y_1}+\alpha \e^{Y_3+Y_1}), \\
\dot{Y}_3 &=(1-\vec{Y}-\mu \e^{Y_1}+\lambda \e^{Y_2}+\alpha \e^{Y_1+Y_2}). 
\end{split}\end{equation}
where $\vec{Y}=\sum_{j=1}^3 \e^{Y_j}$. Note that in these coordinates, the equilibria are at, e.g.\ $Y_1=0$, $Y_2=Y_3=-\infty$. The invariant planes in the $x_j$ coordinates are transformed to $Y_j=-\infty$. However, we are not interested in trajectories which lie in the coordinate planes, only those which are close to them.

\subsection{Addition of feedback}

Pyragas feedback is additive and has the form $F=\Gamma(\vec{x}(t-T)-\vec{x}(T))$
 where $\Gamma$ is a (real) gain matrix and $T$ is the period of the targeted 
periodic orbit. Our choice of coordinates suggests the following slightly altered functional form for the feedback:
\begin{equation}
F=\Gamma\begin{pmatrix} \e^{Y_1(t-T)-Y_1(t)}-1 \\ \e^{Y_2(t-T)-Y_2(t)}-1 \\
\e^{Y_3(t-T)-Y_3(t)}-1 \end{pmatrix}.
\end{equation}
For trajectories close to the periodic orbit, $Y_{j}(t-T)-Y_j(t)\ll 1$ and so the 
feedback terms are approximately of Pyragas form. For this choice of feedback, the 
equilibria and the invariance of the coordinate planes (in the
original $x_j$ coordinates) are preserved. However, we additionally choose
 to use the symmetries of the system to make a further change in the form of the feedback 
which simplifies the subsequent analysis.
The feedback we use is:
\begin{equation}
F=\Gamma\begin{pmatrix} \e^{Y_3(t-\tau)-Y_1(t)}-1 \\ \e^{Y_1(t-\tau)-Y_2(t)}-1 \\
\e^{Y_2(t-\tau)-Y_3(t)}-1 \end{pmatrix},
\end{equation}
where $\tau=T/3$ is one-third of the period of the orbit. Due to the spatiotemporal symmetry of the periodic orbit under the action of $\gamma$, the feedback vanishes at the periodic orbit, and so the periodic orbit is still a solution of the system. However, this feedback does not preserve
 the equilibria or invariant planes (in the original $x_j$ coordinates).

We choose the matrix $\Gamma$ in a similar manner to that in~\cite{PS07b}, as 
follows. We write 
\[\Gamma=EGE^{-1},\] where
\begin{equation}
G=\begin{pmatrix} 0 & 0 & 0 \\ 0 & b_0\cos\beta & -b_0\sin\beta \\
0 & b_0\sin\beta & b_0\cos\beta \end{pmatrix},\quad E=\begin{pmatrix} 1 & -1/2 & \sqrt{3}/2 \\ 1 & -1/2 & -\sqrt{3}/2 \\ 1 & 1 & 0 \end{pmatrix}.
\end{equation}
The matrix $G$ has the form of the feedback matrix used by Fiedler~\etal\cite{Fie06} in a two-dimensional example; stabilising periodic orbits emanating from a subcritical Hopf bifurcation. Recall that the orbit has two unstable 
directions, and one stable direction --- the radial direction. The matrix $E$ is chosen so the feedback is rotated to align with the 
unstable directions, and there is no feedback in the stable direction.

The resulting equations with feedback are
\begin{equation}
\label{eq:GHdelay}
\begin{pmatrix}\dot{Y}_1 \\ \dot{Y}_2 \\  \dot{Y}_3\end{pmatrix}
=I+M\begin{pmatrix} \e^{Y_1} \\ \e^{Y_2} \\ \e^{Y_3} \end{pmatrix} 
+\alpha\begin{pmatrix} \e^{Y_2+Y_3} \\ \e^{Y_3+Y_1} \\ \e^{Y_1+Y_2} \end{pmatrix} 
+\Gamma\begin{pmatrix} \e^{{Y_3}_{\tau}-Y_1}-1 \\e^{{Y_1}_{\tau}-Y_2}-1 \\
e^{{Y_2}_{\tau}-Y_3}-1 \end{pmatrix},
\end{equation}
where 
\[M=\begin{pmatrix} -1 & -\mu-1 & \lambda-1 \\ \lambda-1 & -1 & -\mu-1 \\
 -\mu-1 & \lambda-1 & -1 \end{pmatrix}\qquad\mathrm{and}\qquad {Y_j}_{\tau}=Y_j(t-\tau).
\]

\subsection{Stability analysis}
\label{sec:anal}

We analyse the stability of the periodic orbits close to the heteroclinic cycle in a
 similar manner to the methods used without feedback. We assume we are close to the
 resonant bifurcation, that is, $|1-\delta|\ll 1$, so that the periodic orbit lies close
 to the heteroclinic cycle, and consider the flow close to the periodic orbit. 
The linearised equations close to the equilibrium $\xi_1$ are given by:
\begin{align}
\dot{Y}_2&=\lambda+\Gamma_{22}(Y_{1\tau}-Y_2)+\Gamma_{23}(Y_{2\tau}-Y_3), \label{eq:lin1} \\
\dot{Y}_3&=-\mu+\Gamma_{32}(Y_{1\tau}-Y_2)+\Gamma_{33}(Y_{2\tau}-Y_3), \label{eq:lin2}
\end{align}
where the $\Gamma_{jk}$ are the components of the feedback gain matrix $\Gamma$. As before we neglect the $Y_1$ equation --- since the feedback only acts in directions tangent to the plane containing the periodic orbit, we assume that when trajectories are close enough to the periodic orbit the dynamics in the radial direction are unaffected. That is, near $\xi_1$, the $Y_1$ direction will be contracting and so not affect the stability of the orbit. In section~\ref{sec:just} we show numerical results which support this assumption.

Recall that the periodic orbits we are attempting to stabilise are spatiotemporally symmetric under
the action of $\gamma$. We make use of this in the following. At each
equilibrium, we define a contracting direction, and an expanding
direction. At $\xi_j$, the contracting direction $Y_c$ is the $Y_{j-1}$
direction, and the expanding direction $Y_e$ is the $Y_{j+1}$ direction. 

Unlike in the case without feedback, we cannot solve the linear
equations explicitly, and so we make the following approximations.
Let $Y^{\star}(t;\mu,\lambda)$ be the periodic orbit for the original system (in the logarithmic coordinates). Then $Y^{\star}$ is still a solution of the system with feedback. Consider solving the delay differential equation for the system with feedback for a trajectory $Y(t)$ which starts close to $Y^{\star}$. That is, for $-\tau<t<0$, $Y(t)$ is close to $Y^{\star}(t)$. Then for $0<t<\tau$, the feedback terms in the delay differential equation will be small, that is, the equations will only be a small perturbation from the original system. By continuity, the solution $Y(t)$ for $0<t<\tau$ will also be close to $Y^{\star}$.

Set $t=0$ as the trajectory intersects the plane $\Hin_{j}$ ($Y_{j-1}=H$,  for $H=\log h$, $0< h\ll 1$) on the $i$th time the trajectory passes close to an equilibrium, $\xi_j$. 
 With no feedback, the local part of the trajectory can be written down exactly. The expanding and contracting components, $Y_e(t)$ and $Y_c(t)$ satisfy:
\begin{align} 
Y_e(t)& =Y_e^i+\lambda t,  \\
Y_c(t)&=H-\mu t,
\end{align}
for $t\in[0,T_i)$, where $T_i$ is the length of time spent near the
equilibrium (\ie in the small box) 
and $Y_e^i$ is the expanding coordinate of the trajectory as it intersects the plane $Y_c=H$.
 
\begin{figure}
\psfrag{t0}{$t=0$}
\psfrag{xc}{$x_c$}
\psfrag{xe}{$x_e$}
\psfrag{Hin}{$\Hin_j$}
\psfrag{Hout}{$\Hout_j$}
\psfrag{xi1}{$\xi_j$}
\psfrag{tti}{$t=T_i$}
\psfrag{mtau}{$t=-\tau$}
\psfrag{Tim1}{$t=-T_{i-1}$}
\psfrag{Ti}{$t=T_i$}
\psfrag{i}{$i$}
\psfrag{im1}{$i-1$}
\begin{center}
\epsfig{figure=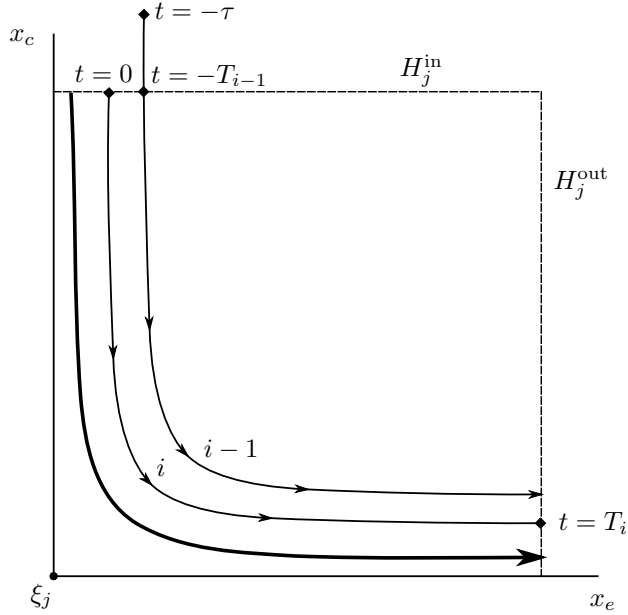,height=8cm}
\caption{\label{fig:lm}The figure shows a schematic of the local flow as the trajectory passes an equilibrium $\xi_j$ for the $(i-1)$th and $i$th time. The bold line indicates the periodic orbit. We set $t=0$ as the trajectory intersects $\Hin_j$ on the $i$th pass. The trajectory intersects $\Hin_j$ at $t=-T_{i-1}$ on the $(i-1)$th pass, which does not coincide with $t=-\tau$. Note that here we show the $(i-1)$th trajectory passing the same equilibrium as the $i$th trajectory --- in actuality the $(i-1)$th pass is of the previous equilibrium, but because of the symmetries in the system this is equivalent, and schematically simpler to show in the figure. The diamonds ($\blacklozenge$) indicate the points of the trajectory at which times are indicated on the figure. The dashed lines indicate the surfaces $\Hin_j$ and $\Hout_j$.}
\end{center}
\end{figure}

For the system with feedback, we cannot explicitly solve the linearised equations. Given the argument above, we assume that we start sufficiently close to the periodic orbit that solutions are only a small perturbation away from those for the case with no feedback. That is, we write, for $t\in[0,T_i)$, 
\begin{align} 
Y_e(t)& =Y_e^i+\lambda t+f_i(t), \label{eq:untrunc1} \\
Y_c(t)&=H-\mu t+g_i(t), \label{eq:untrunc2}
\end{align}
where $f_i(t)$ and $g_i(t)$ are functions which satisfy
\[f_i(t),g_i(t)\ll 1,\qquad f_i(0)=g_i(0)=0,\]
and if the trajectory is exactly the periodic orbit, $f_i(t),g_i(t)\equiv 0$. We will use this assumed form of the local flow together with equations~\eqref{eq:lin1} and~\eqref{eq:lin2} and the global flow as before to derive a new return map. This gives recurrence relations for $Y_e^i$, and the functions $f_i(t)$ and $g_i(t)$. Figure~\ref{fig:lm} shows a schematic of the local flow past an equilibrium $\xi_j$, in the original $x_j$ coordinates.

We again ignore the time the trajectory spends near the heteroclinic
connections but away from the  
equilibria, so for $t\in[-T_{i-1},0)$, the flow is given by
\begin{align} 
Y_e(t)& =Y_e^{i-1}+\lambda t+f_{i-1}(t), \label{eq:linm1} \\
Y_c(t)&=H-\mu t+g_{i-1}(t), \label{eq:linm2}
\end{align}
Using the symmetry $\gamma$, we can rewrite the linear
equations~\eqref{eq:lin1} and~\eqref{eq:lin2} as
\begin{align}
\dot{Y}_e&=\lambda+\Gamma_{22}(Y_{e\tau}-Y_e)+\Gamma_{23}(Y_{c\tau}-Y_c), \label{eq:lin_e} \\
\dot{Y}_c&=-\mu+\Gamma_{32}(Y_{e\tau}-Y_e)+\Gamma_{33}(Y_{c\tau}-Y_c), \label{eq:lin_c}
\end{align}
where the delayed terms are the corresponding coordinates near the previous equilibrium, that is
\begin{align}
Y_{e\tau}\equiv{Y_e}(t-\tau)&=Y_e(-\tau)+\lambda t+f_{i-1}(t) \\
Y_{c\tau}\equiv{Y_c}(t-\tau)&=Y_c(-\tau)-\mu t +g_{i-1}(t)
\end{align}

The time of flight of the trajectory between the planes $\Hin_j$ and $\Hout_j$,
 $T_{i-1}$, will not be equal to the delay time $\tau$ except when the trajectory is
 exactly on the periodic orbit (see figure~\ref{fig:lm}). In order to find the coordinates at
 $t=-\tau$, we assume the flow given by~\eqref{eq:linm1}
 and~\eqref{eq:linm2} is also valid for $t\in(-\tau,-T_{i-1}]$. The coordinates of the 
trajectory at $t=-\tau$ are therefore:
\begin{align}
Y_e(-\tau) &=Y_e^{i-1}+\lambda(T_{i-1}-\tau)+f_{i-1}(T_{i-1}-\tau), \label{eq:y2ti} \\
Y_c(-\tau) &=H-\mu(T_{i-1}-\tau)+g_{i-1}(T_{i-1}-\tau).\label{eq:y3ti}
\end{align}
Writing $T_i=\tau+\Delta_i$, where $\Delta_i/\tau\ll 1$ (since we are
close to the periodic orbit) and expanding $f_{i-1}$ and $g_{i-1}$
about zero gives 
\begin{align}
f_{i-1}(\Delta_i)&=f_{i-1}(0)+\Delta_{i-1}f_{i-1}'(0)+\dots\approx \Delta_{i-1}f_{i-1}'(0), \\
g_{i-1}(\Delta_i)&=g_{i-1}(0)+\Delta_{i-1}g_{i-1}'(0)+\dots\approx \Delta_{i-1}g_{i-1}'(0),
\end{align}
since $f_i(0)=g_i(0)=0$.
Substituting into~\eqref{eq:y2ti} and~\eqref{eq:y3ti} gives:
\begin{align}
Y_e(-\tau)&=Y_e^{i-1}+\Delta_{i-1}(\lambda+f_{i-1}'(0)), \label{eq:delay1}\\
Y_c(-\tau) &=H+\Delta_{i-1}(-\mu+g_{i-1}'(0)). \label{eq:delay2}
\end{align}

We also have that for $t\in[0,T_i)$,
\[\dot{Y}_e(t)=\lambda+f_i'(t),\quad\dot{Y}_c(t)=-\mu+g_i'(t)\]
Substituting~\eqref{eq:untrunc1},~\eqref{eq:untrunc2},~\eqref{eq:delay1} and~\eqref{eq:delay2} into equations~\eqref{eq:lin_e} and~\eqref{eq:lin_c} we find:
\begin{multline}f_i'(t)=\Gamma_{22}[Y_e^{i-1}-Y_e^i+\Delta_{i-1}(\lambda+f_{i-1}'(0))+f_{i-1}(t)-f_i(t)]
+ \\ \Gamma_{23}[\Delta_{i-1}(-\mu+g_{i-1}'(0))+g_{i-1}(t)-g_i(t)],
 \end{multline}
\begin{multline}
g_i'(t)=\Gamma_{32}[Y_e^{i-1}-Y_e^i+\Delta_{i-1}(\lambda+f_{i-1}'(0))+f_{i-1}(t)-f_i(t)]
+ \\ \Gamma_{33}[\Delta_{i-1}(-\mu+g_{i-1}'(0))+g_{i-1}(t)-g_i(t)]. \end{multline}
These expressions are true for all $t\in[0,T_i)$, so we set $t=0$ to simplify and find:
\begin{align}
f_i'(0)=&\Gamma_{22}[Y_e^{i-1}-Y_e^i+\Delta_{i-1}(\lambda+f_{i-1}'(0))]
+\Gamma_{23}[\Delta_{i-1}(-\mu+g_{i-1}'(0))],  \\
g_i'(0)=&\Gamma_{32}[Y_e^{i-1}-Y_e^i+\Delta_{i-1}(\lambda+f_{i-1}'(0))]
+\Gamma_{33}[\Delta_{i-1}(-\mu+g_{i-1}'(0))], 
\end{align}
that is, a recurrence relation for $f_i'(0)$
and $g_i'(0)$ if the $\Delta_i$ and $Y_e^i$ are known. We write $X_i=Y_e^i-H$, $\lambda_i=f_i'(0)+\lambda$ and $\mu_i=g_i'(0)-\mu$ to further simplify:
\begin{align}
\lambda_i=&\lambda+\Gamma_{22}(X_{i-1}-X_i+\Delta_{i-1}\lambda_{i-1})
+\Gamma_{23}\Delta_{i-1}\mu_{i-1}, \label{eq:rr_a_1} \\
\mu_i=&-\mu+\Gamma_{32}(X_{i-1}-X_i+\Delta_{i-1}\lambda_{i-1})
+\Gamma_{33}\Delta_{i-1}\mu_{i-1}, \label{eq:rr_a_2}
\end{align}

We next find an expression for $\Delta_i$ which we use to find a recurrence relation for the $X_i$. Recall that $Y_e(T_i)=H$, so from~\eqref{eq:untrunc1} we have
\begin{align}
H=Y_e(T_i)& =Y_e^i+\lambda T_i+f_i(T_i), \\
X_i&=-\lambda T_i-f_i(T_i).
\end{align}

In order to be able to get tractable results in what follows, we need to invert the above equation for $T_i$. Motivated by numerical results, which we give in section~\ref{sec:just}, we make the following assumption:
\[f_i(T_i)\approx T_if_i'(0),\]
that is, that $f_i$ is approximately a linear function of $t$. 
Using this gives us
\[T_i\approx-\frac{X_i}{\lambda_i}\]
so 
\begin{equation}\label{eq:Deltai}
\Delta_i=T_i-\tau\approx-\frac{X_i}{\lambda_i}+\frac{X^{\star}}{\lambda}\end{equation}
where $X^{\star}=-\lambda\tau=\log A/(1-\delta)$.

We make a similar assumption on the $g_i$, that is, $g_i(T_i)\approx T_ig_i'(0)$, and then use~\eqref{eq:untrunc2} to find $Y_c(T_i)$:
\[Y_c^o\equiv Y_c(T_i)\approx H+(-\mu+g_i'(\tau))T_i=H-\frac{\mu_i}{\lambda_i}X_i,\]
which is an expression for the local map $\phi_{\mathrm{loc}}:Y_e^i\rar Y_c^o$. We assume that the global map $\phi_{\mathrm{far}}$ is of the same form as the case without feedback~\eqref{eq:far} when we are close enough to the periodic orbit, and hence find a return map for the $X_i$:
\[X_{i+1}=\log A-\frac{\mu_i}{\lambda_i}X_i.\]

Substituting equation~\eqref{eq:Deltai} into equations~\eqref{eq:rr_a_1} and~\eqref{eq:rr_a_2} results in a third order recurrence system:
\begin{equation}
\label{eq:rr}
\begin{split}
X_i&=\log A-\frac{\mu_{i-1}}{\lambda_{i-1}}X_{i-1}, \\
\lambda_i&=\lambda+\Gamma_{22}\left(\frac{\lambda_{i-1}}{\lambda}X^{\star}+\frac{\mu_{i-1}}{\lambda_{i-1}}X_{i-1}\right)+
\Gamma_{23}\mu_{i-1}\left(\frac{X^{\star}}{\lambda}-\frac{X_{i-1}}{\lambda_{i-1}}\right) -\Gamma_{22}\log A, \\
\mu_i&=-\mu+\Gamma_{32}\left(\frac{\lambda_{i-1}}{\lambda}X^{\star}+\frac{\mu_{i-1}}{\lambda_{i-1}}X_{i-1}\right)+
\Gamma_{33}\mu_{i-1}\left(\frac{X^{\star}}{\lambda}-\frac{X_{i-1}}{\lambda_{i-1}}\right) -\Gamma_{32}\log A.
\end{split}
\end{equation} 
Note that when
$\Gamma_{jk}\equiv 0$, the recurrence relation reduces to that for the
system with no feedback, as expected. 
This system of three recurrence relations has a fixed point at 
\[\lambda_i=\lambda,\ \mu_i=-\mu,\ X_i=X^{\star}\]
which corresponds to the periodic orbit in the flow. The stability of
the fixed point in the recurrence relation will correspond to the
stability of the periodic orbit in the flow.

The Jacobian matrix $J$ of~\eqref{eq:rr} at this fixed point is:
\[J=\begin{pmatrix} \delta &  -\delta \hat{X}&  -\hat{X} \\
-\delta(\Gamma_{22}-\Gamma_{23}) & \hat{X}[\Gamma_{22}+\delta(\Gamma_{22}-\Gamma_{23})]& 
\Gamma_{22}\hat{X} \\
-\delta(\Gamma_{32}-\Gamma_{33}) & \hat{X}[\Gamma_{32}+\delta(\Gamma_{32}-\Gamma_{33})]& 
\Gamma_{32}\hat{X} \end{pmatrix}\]
where $\hat{X}=X^{\star}/\lambda$. The characteristic equation of $J$ is 
\begin{equation}\label{eq:cp}
m^3+am^2+bm+c=0,\end{equation}
where
\begin{align}
a&= \hat{X}\delta(\Gamma_{23}-\Gamma_{22})-\hat{X}(\Gamma_{22}+\Gamma_{32})-\delta, \\
b&= {\hat{X}}^2\delta(\Gamma_{22}\Gamma_{33}-\Gamma_{23}\Gamma_{32})+\hat{X}\delta(\Gamma_{33}+\Gamma_{22}), \\
c&= -{\hat{X}}^2\delta(\Gamma_{22}\Gamma_{33}-\Gamma_{23}\Gamma_{32}).
\end{align}
The fixed point will be unstable if~\eqref{eq:cp} has any solutions with $|m|>1$, so curves with $|m|=1$ define stability boundaries of the periodic orbit. 
 Recall that $\Gamma=EGE^{-1}$ and is a
function of just two parameters, $b_0$ and $\beta$.
We consider the stability of the periodic orbit as the parameters
$\delta$ and $b_0$ are varied. 

\subsection{Determination of stability boundaries}

We split our investigation of the stability boundaries into three cases.
We introduce the bifurcation parameter $\nu=\delta-1$.  Without feedback, the heteroclinic cycle is stable in $\nu>0$ and the periodic orbits exist and are unstable in $\nu>0$. We consider
analytically the limits of the stability boundary curves as $\nu, b_0\rar
0$. The boundaries can actually be computed exactly (although the algebra is rather nasty), since the eigenvalues are the roots of a cubic. We plot the boundaries for specific parameter values in figure~\ref{fig:stab_anal}.
In section~\ref{sec:num} we compute the stability of the periodic orbit in the original system~\eqref{eq:GHdelay}, numerically using the continuation package \textsc{dde-biftool}.

\subsubsection*{Case 1: $m=1$}

A stability boundary with $m=1$ corresponds to a steady state
bifurcation of the periodic orbit. This occurs when $1+a+b+c=0$, that is
\[1-\delta+\hat{X}\delta(\Gamma_{23}+\Gamma_{33})-\hat{X}(\Gamma_{22}+\Gamma_{32})=0.\]
It can easily be computed that 
\begin{align}
\Gamma_{23}+\Gamma_{33}&=\frac{b_0}{3}(\cos\beta-\sqrt{3}\sin\beta) \\
\intertext{and}
\Gamma_{22}+\Gamma_{32}&=\frac{b_0}{3}(\cos\beta+\sqrt{3}\sin\beta)
\end{align}
In the limit  $\nu, b_0\rar 0$, using $\hat{X}=-\frac{\log A}{\lambda\nu}$, we find
\begin{equation}\label{eq:m1}
\nu^2= p_1b_0,\qquad p_1=\frac{2\log A\sin\beta}{\sqrt{3}\lambda}.\end{equation}
It is also simple to calculate that in the limit $\nu\rar 0$ the eigenvalue which goes through $m=1$ as this curve is crossed is greater than $1$ if $\nu^2>p_1b_0$ and less than $1$ if $\nu^2<p_1b_0$.

\subsubsection*{Case 2: $m=-1$}
A stability boundary with $m=-1$ will correspond to a period-doubling bifurcation of the periodic orbit. These curves will have $-1+a-b+c=0$, that is,
\[-1-\delta+\hat{X}\delta(\Gamma_{23}-2\Gamma_{22}-\Gamma_{33})-
\hat{X}(\Gamma_{22}+\Gamma_{32})-
2{\hat{X}}^2\delta(\Gamma_{22}\Gamma_{33}-\Gamma_{23}\Gamma_{32})=0\]
Again, we can compute the coefficients
\begin{align}
\Gamma_{22}\Gamma_{33}-\Gamma_{23}\Gamma_{32}&=\frac{b_0^2}{3}, \\
\Gamma_{23}-2\Gamma_{22}-\Gamma_{33}&=-\frac{b_0}{3}(7\cos\beta+\sqrt{3}\sin
\beta),
\end{align}
and in the same limit as above, we find 
\begin{equation}\label{eq:m2}1-\frac{\log A}{3\lambda}(4\cos\beta+\sqrt{3}\sin\beta)\frac{b_0}{\nu} + \frac{(\log A)^2}{3\lambda^2}\frac{b_0^2}{\nu^2}=0\end{equation}
so there are two solutions $b_0=c_{\pm}\nu$ for some $c_{\pm}$ function of $A$, $\lambda$ and $\beta$. The direction of the bifurcation as these lines are crossed in this case depends on $\beta$.

\subsubsection*{Case 3: $m=\e^{i\theta}$, $\theta\neq n\pi$}
For $m=\e^{i\theta}$, $\theta\neq n\pi$, it can easily be computed that we must have $b+c^2-ac=1$.
 The computations in this case are messier, so we omit them, and give the resulting curve in the limit $b_0,\nu\rightarrow 0$,
\begin{equation}\label{eq:m3} \left(\frac{b_0}{\nu}\right)^4\frac{(\log A)^4}{3\lambda^4}-
\left(\frac{b_0}{\nu}\right)^3\frac{2(\log A)^3}{3\lambda^3}(2\cos\beta+\sqrt{3}\sin\beta)-
\frac{b_0^2}{\nu}\frac{(\log A)^2}{\lambda^2}-\frac{b_0}{\nu}\frac{4\log A}{\lambda}\cos\beta =3\end{equation}
The direction of the bifurcation again will depend on $\beta$.

\begin{figure}
\psfrag{mu}{$\mu$}
\psfrag{b0}{$b_0$}
\begin{center}
\epsfig{figure=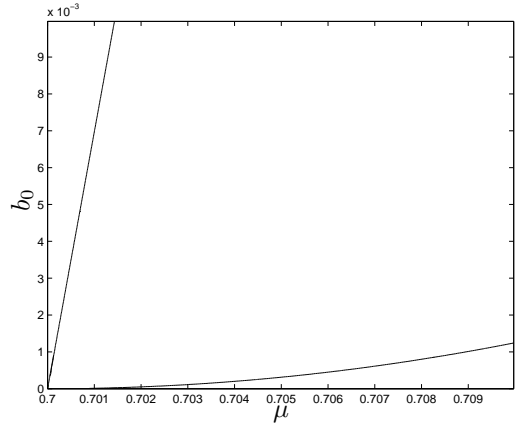,height=6cm}
\caption{The figure shows the stability boundaries of the periodic orbit, that is, curves of solutions $|m|=1$ to equation~\eqref{eq:cp}, as $\mu$ and $b_0$ are varied. The lower boundary is a steady state bifurcation ($m=1$) and the left hand boundary is a period-doubling bifurcation ($m=-1$). The orbit is stable in the wedge between these lines. Remaining parameters are $\lambda=0.7$, $\beta=\pi/4$, $\alpha=0.01$, so $\nu=0$ at $\mu=0.7$ (left hand side of figure).}
\label{fig:stab_anal}
\end{center}
\end{figure}

The curves~\eqref{eq:m1},~\eqref{eq:m2} and~\eqref{eq:m3} describe the limiting cases of the stability boundaries of the periodic orbit as the point $\nu=b_0=0$ is approached. Since the characteristic polynomial~\eqref{eq:cp} is cubic, it can be solved for any values of $\nu$ and $b_0$. In figure~\ref{fig:stab_anal} we plot the solutions $|m|=1$ of~\eqref{eq:cp} for a specific set of parameter values. In this case only the curves corresponding to $m=1$ and $m=-1$ are stability boundaries. The lower boundary is the quadratic curve for $m=1$ and the left hand boundary is a straight line corresponding to $m=-1$. The remaining curves with $|m|=1$ do not form stability boundaries in this case because the periodic orbit is already unstable in the regions in which they exist.

We can see that for these parameter values, the periodic orbit is stable for a wide range of parameters, and specifically, can be stabilised arbitrarily close to the heteroclinic cycle, that is, for arbitrarily large period. That is, for any $\nu>0$ we can find a $b_0$ for which the periodic orbit is stable. In fact, for this particular case, we see that as $\nu$ gets smaller, in order for the orbit to be stable, we have to choose $b_0$, the gain parameter, to be increasingly small. This seems a rather surprising result - that as the period of the targeted orbit increases, the amplitude of the gain parameter tends towards zero.

\subsection{Stability of the heteroclinic cycle}

We note that the recurrence relations~\eqref{eq:rr} have a second solution, $\lambda_i=\lambda$, 
$\mu_i=-\mu$, $X_i\rar -\infty$, which corresponds to the heteroclinic cycle. We 
can consider the stability of this solution by considering the solutions
 to~\eqref{eq:cp} in the limit $\hat{X}\rar -\infty$. In this limit,
 $a\sim O(\hat{X})$, $b\sim O(\hat{X}^2)$ and  $c\sim O(\hat{X}^2)$. The cubic equation~\eqref{eq:cp} therefore has one solution with $m\sim O(1)$ and two solutions with $m\sim O(\hat{X})$. Therefore this fixed point is always unstable in the recurrence relation, and so the heteroclinic cycle is always unstable in the flow.

\section{Numerical results}
\label{sec:num}

We use the Matlab package \textsc{dde-biftool}~\cite{ddebiftool} to numerically analyse the stability of periodic orbits in the system~\eqref{eq:GHdelay}. The delay time $\tau$ was set equal to the period of the bifurcating periodic orbits (and so is a function of $\mu$), and was calculated numerically from the system with no feedback. Parameters used were the same as those used to produce figure~\ref{fig:stab_anal}.

\begin{figure}
\psfrag{mu}{$\mu$}
\psfrag{b0}{$b_0$}
\psfrag{lam}{}
\begin{center}
\epsfig{figure=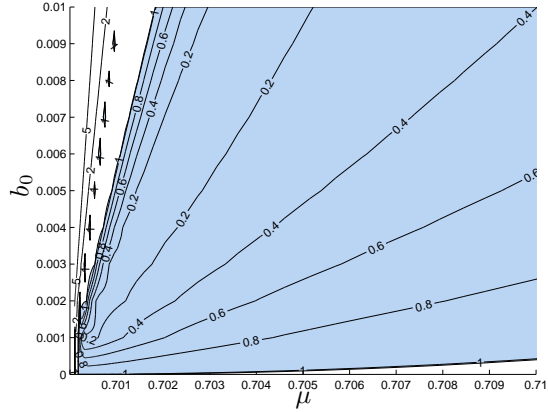,height=6cm}
\caption{The figure shows a contour plot of the magnitude of the largest Floquet
multipliers of the periodic orbit, as the parameters $b_0$ and $\mu$ are varied. The orbit is stable when the largest Floquet multiplier has magnitude less than one, which is indicated by the shading. Remaining parameter
values are $\lambda=0.7$, $\beta=\pi/4$, $\alpha=0.01$, so $\nu=0$ at $\mu=0.7$ (left hand side of figure). This figure was produced using \textsc{dde-biftool}.}
\label{fig:stab1}
\end{center}
\end{figure}

Figure~\ref{fig:stab1} shows a contour plot of the amplitude of the largest Floquet multiplier as the parameters $b_0$ and $\mu$ are varied. The periodic orbit is stable when all Floquet multipliers have amplitude less than $1$, and this region is indicated by the shading in figure~\ref{fig:stab1}. Comparison with figure~\ref{fig:stab_anal}, showing the stability as calculated analytically, shows a very good agreement between the location of the stability boundaries. The shapes of the boundaries also agrees, that is, the left hand boundary is a straight line, whereas the lower boundary is part of a parabola.
The nature of the bifurcations that occur as the boundaries are crossed also agrees with the analytical result. That is, the left hand boundary is a period-doubling bifurcation, with a critical Floquet multiplier equal to $-1$, and the lower boundary is a steady state bifurcation with a critical Floquet multiplier of $+1$.

\begin{figure}
\psfrag{t}{$t$}
\psfrag{y}{$Y_j$}
\psfrag{dy}{\raisebox{0.2cm}{\hspace{-0.2cm}$\frac{\mathrm{d}Y_j}{\mathrm{d}t}$}}
\psfrag{yd}{\raisebox{0.1cm}{\hspace{-1.3cm}$Y_{j-1}(t-\tau)-Y_j(t)$}}
\begin{center}
\subfigure[\label{fig:int1}]{\epsfig{figure=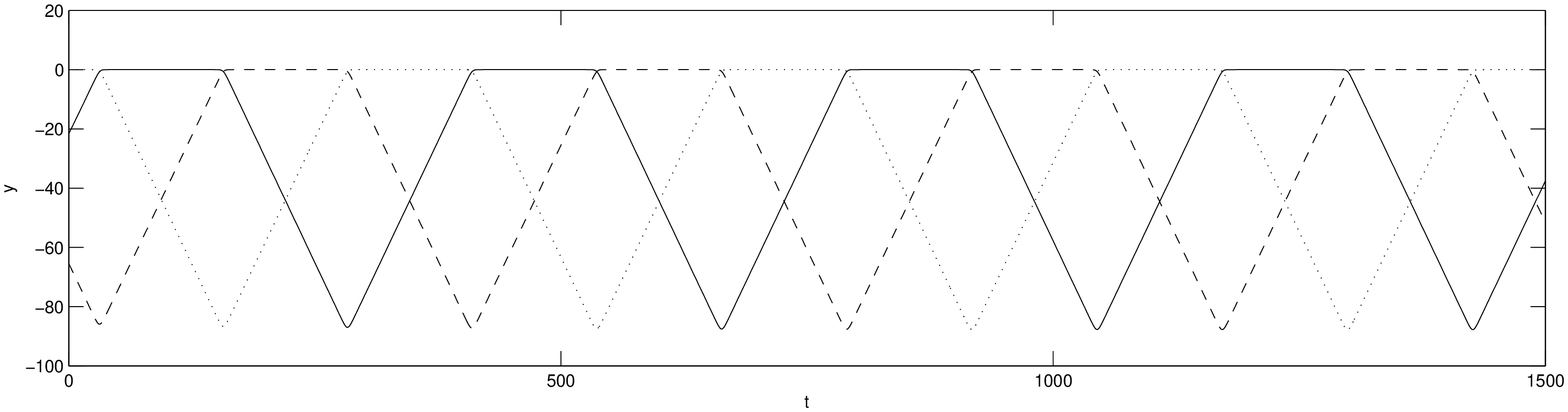,width=0.9\textwidth}}
\subfigure[\label{fig:intderiv1}]{\epsfig{figure=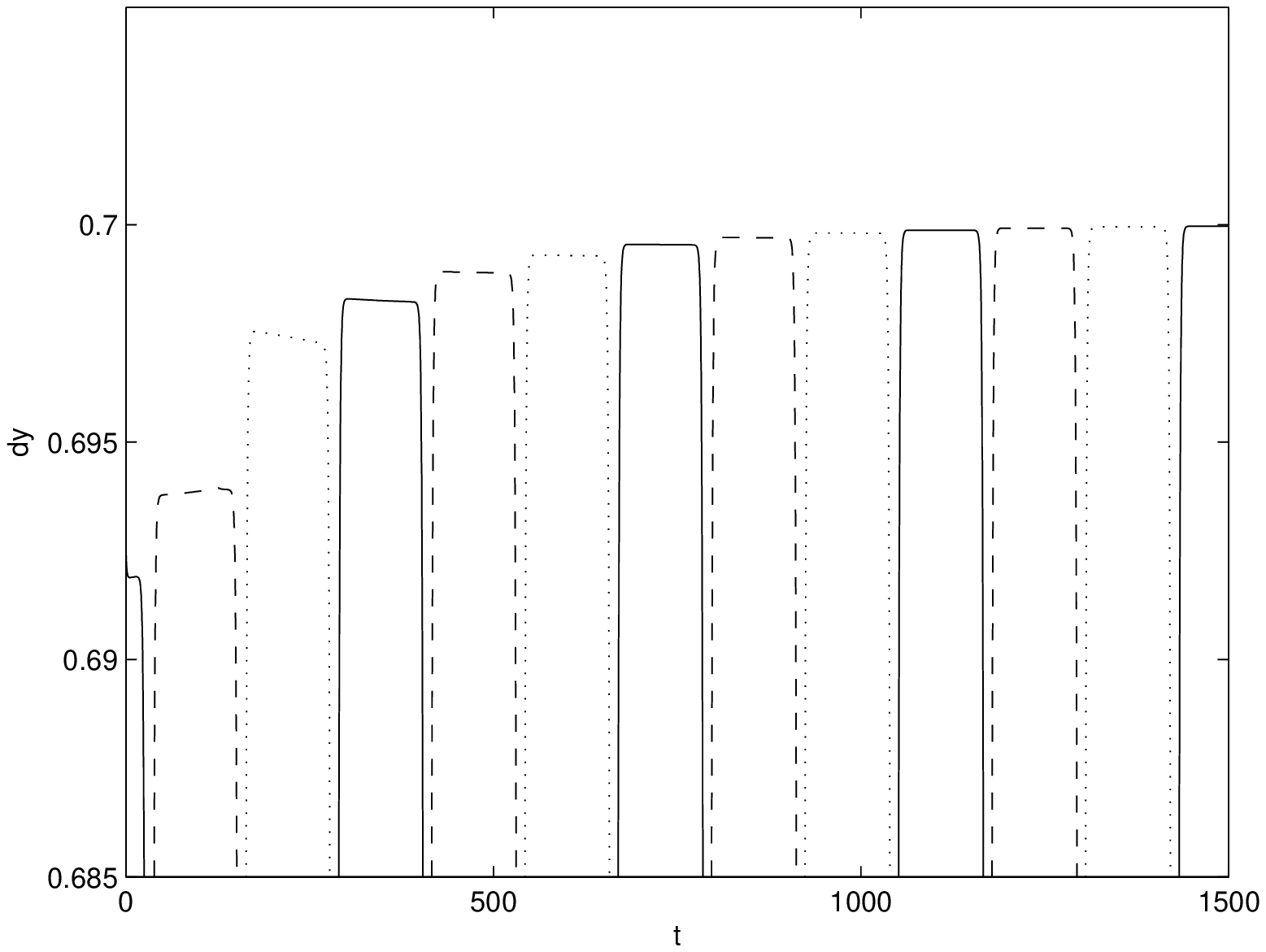,width=0.45\textwidth}}
\subfigure[\label{fig:intdiff1}]{\epsfig{figure=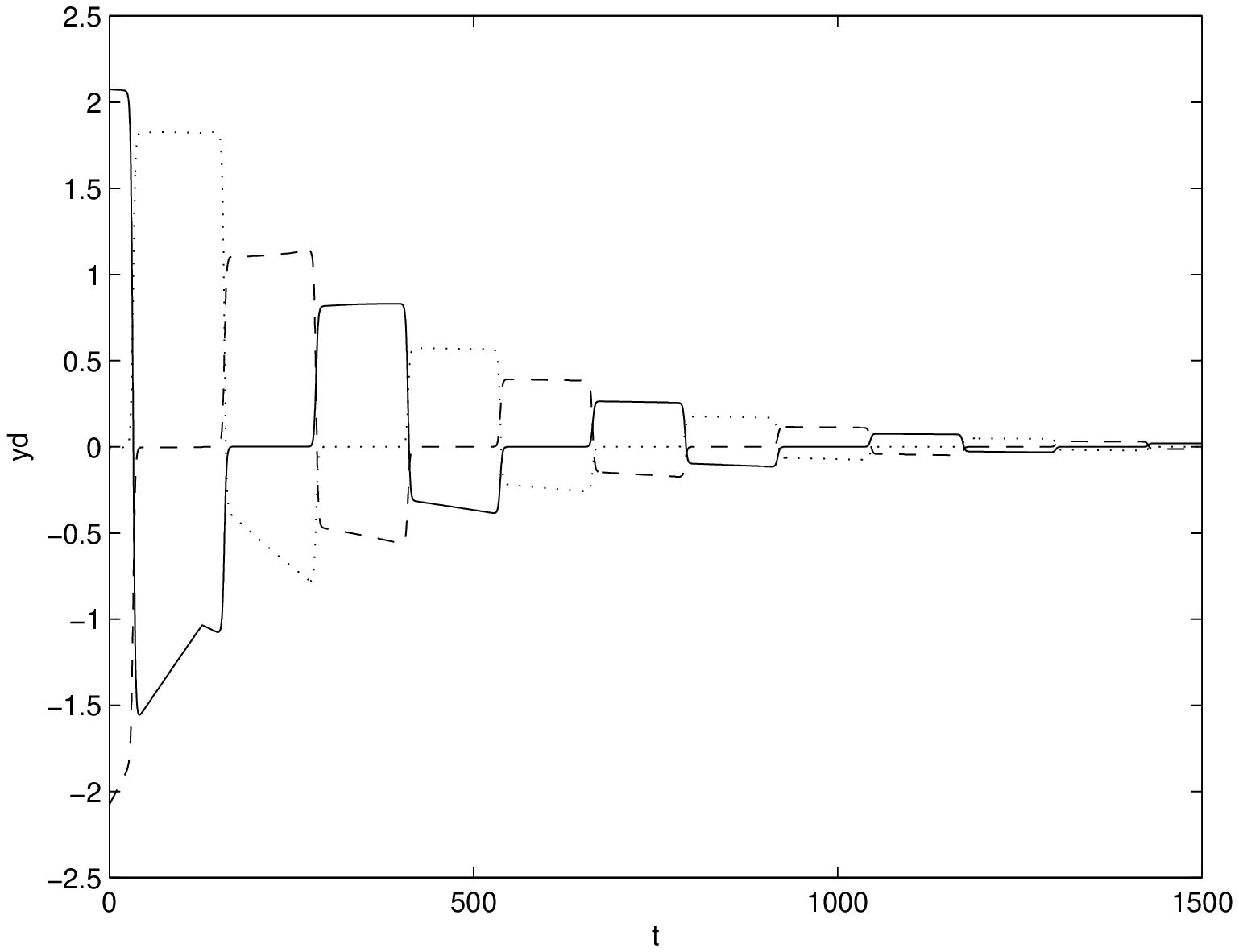,width=0.45\textwidth}}
\caption{\label{fig:int}Figure (a) shows forward integration of 
equations~\eqref{eq:GHdelay}, using Matlab routine \texttt{dde23}. Figure (b) shows the derivative of the curves shown in
 (a), it can be seen that the derivative of the expanding direction at each equilibria
 tends to $0.7$ as the periodic orbit is approached. Figure (c) shows the feedback
 terms, $Y_{j-1}(t-\tau)-Y_j(t)$, which clearly tends to zero as the orbit is 
approached. We can also see that the feedback term corresponding to the coordinates 
$Y_{j-1}(t-\tau)-Y_j(t)$ when we are near equilibria $\xi_i$ is much smaller than the 
other feedback terms. Parameters are $\mu=0.7012$, $\lambda=0.7$, $b_0=0.0015$, $\beta=\pi/4$,
 $\alpha=0.01$}
\end{center}
\end{figure}

Forward integration of the equations~\eqref{eq:GHdelay} also confirms the stability
results. In figure~\ref{fig:int} we show results from such an integration. 
We also show the derivative of the coordinates and the feedback terms. It can be seen that as the periodic orbit is approached, the derivative of the expanding coordinate tends to $\lambda$ (in this case, $\lambda=0.7$), and the feedback terms tend towards zero.

\subsection{Justification of assumptions}
\label{sec:just}

In section~\ref{sec:anal} we make a number of assumptions regarding the form of solutions to the delay differential equations. Firstly, we assume that the `radial' direction does not affect the stability of the periodic orbits, and so we neglect this coordinate in our construction of a Poincar\'e map. Secondly, that trajectories starting near the periodic orbits will be only small perturbations from the form of solutions to the original equations without feedback. Thirdly, we make the assumption that  $f_i(T_i)\approx T_if_i'(0)$, and $g_i(T_i)\approx T_ig_i'(0)$.

Here, we address each assumption in turn and show that our numerical results support these assumptions.

Figure~\ref{fig:intdiff1} shows the feedback terms in a forward integration of equations~\eqref{eq:GHdelay} as the periodic orbit is approached. It can be seen here that the feedback terms corresponding to the radial direction are much smaller than the other feedback terms --- on this scale they cannot be distinguished from zero. Hence the affect of the feedback on the radial direction is negligible and this assumption is justified.

\begin{figure}
\psfrag{t}{\raisebox{-0.2cm}{$t$}}
\psfrag{yj}{\raisebox{0.1cm}{$Y_j$}}
\psfrag{ydiff}{\raisebox{0.1cm}{\hspace{-1.3cm}$Y_{j-1}(t-\tau)-Y_j(t)$}}
\begin{center}
\subfigure[\label{fig:unstable1}]{\epsfig{figure=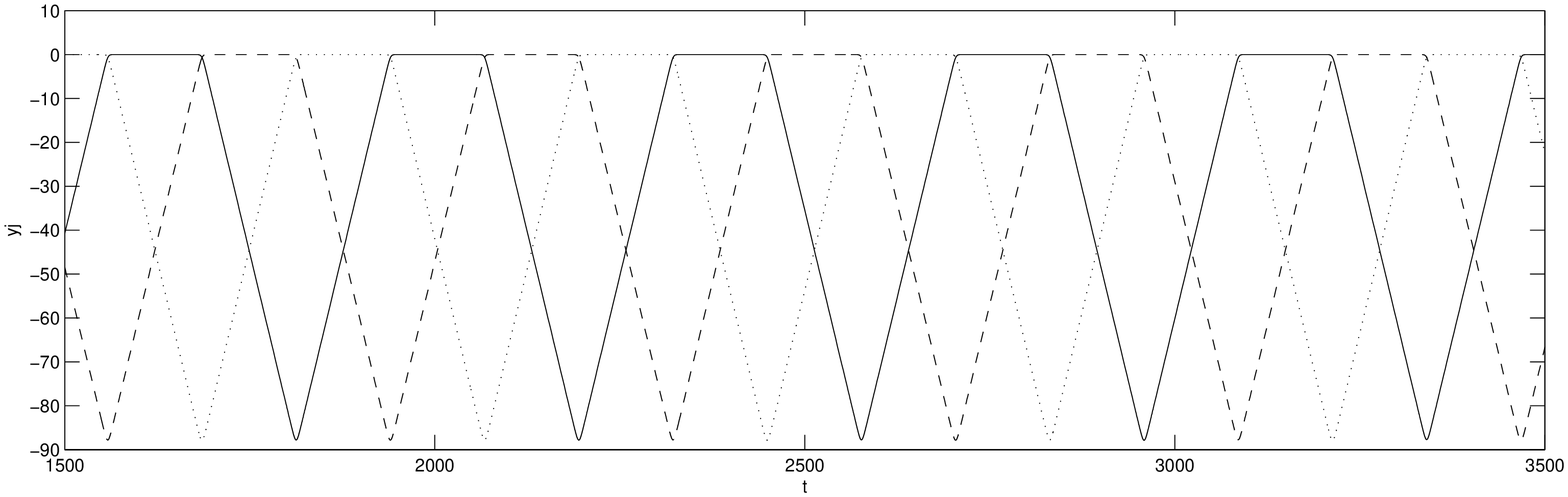,width=0.9\textwidth}}
\subfigure[\label{fig:unstable2}]{\epsfig{figure=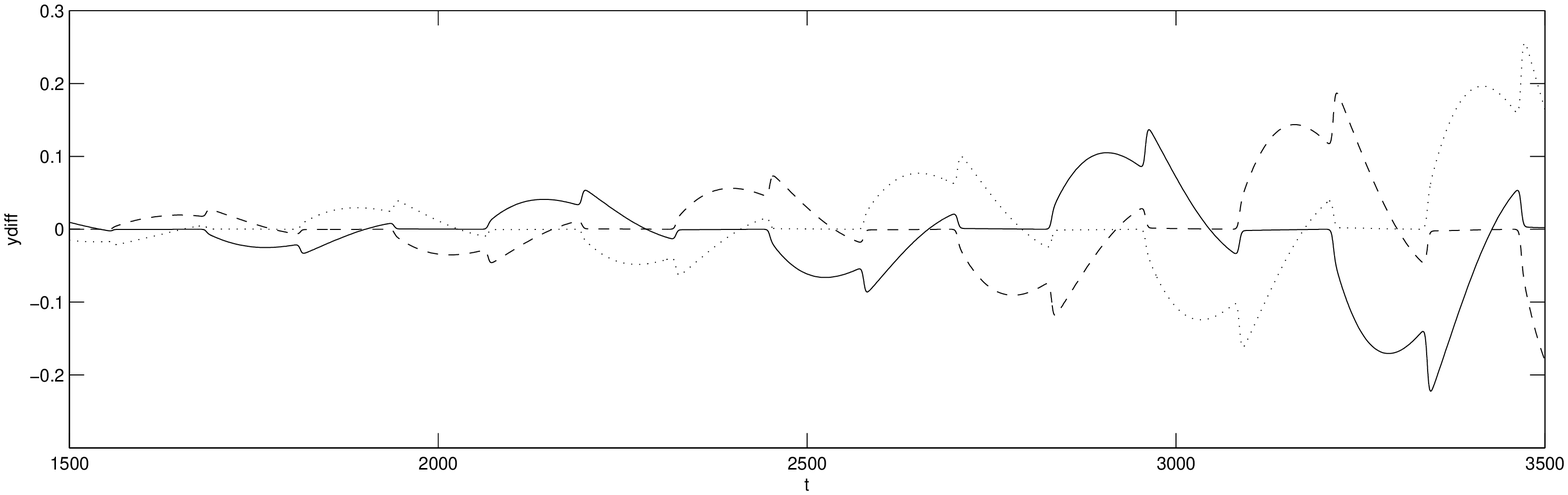,width=0.9\textwidth}}
\caption{\label{fig:int_un}The figure shows forward integration of 
equations~\eqref{eq:GHdelay}, using Matlab routine \texttt{dde23}, for a parameter set in which the periodic orbit is unstable. Figure (a) shows a time series of the trajectory and figure (b) shows the feedback
 terms, $Y_{j-1}(t-\tau)-Y_j(t)$. Parameters are $\mu=0.7012$, $\lambda=0.7$, $b_0=0.008$, $\beta=\pi/4$,
 $\alpha=0.01$}
\end{center}
\end{figure}

Regarding the second assumption, it can be clearly seen in figure~\ref{fig:intdiff1} that the feedback terms decay to zero as the periodic orbit is approached. However, this is to be expected in the case that the periodic orbit is stable. In figure~\ref{fig:int_un} we show the results of an integration in which the periodic orbit is unstable. It can be seen from the time series in~\ref{fig:unstable1} that the trajectories still remain approximately of the form of the periodic orbit even though the trajectory is moving away. Figure~\ref{fig:unstable2} shows the feedback terms, a measure of how close the trajectory is to the periodic orbit. Although they are increasing in magnitude, they do so in the same manner one would expect for an unstable periodic orbit in ordinary differential  equations. That is, by starting trajectories close enough to the periodic orbit, the feedback magnitude can be bounded above for arbitrarily long time.

\begin{figure}
\psfrag{j}{\raisebox{-0.1cm}{$i$}}
\begin{center}
\subfigure[]{\epsfig{figure=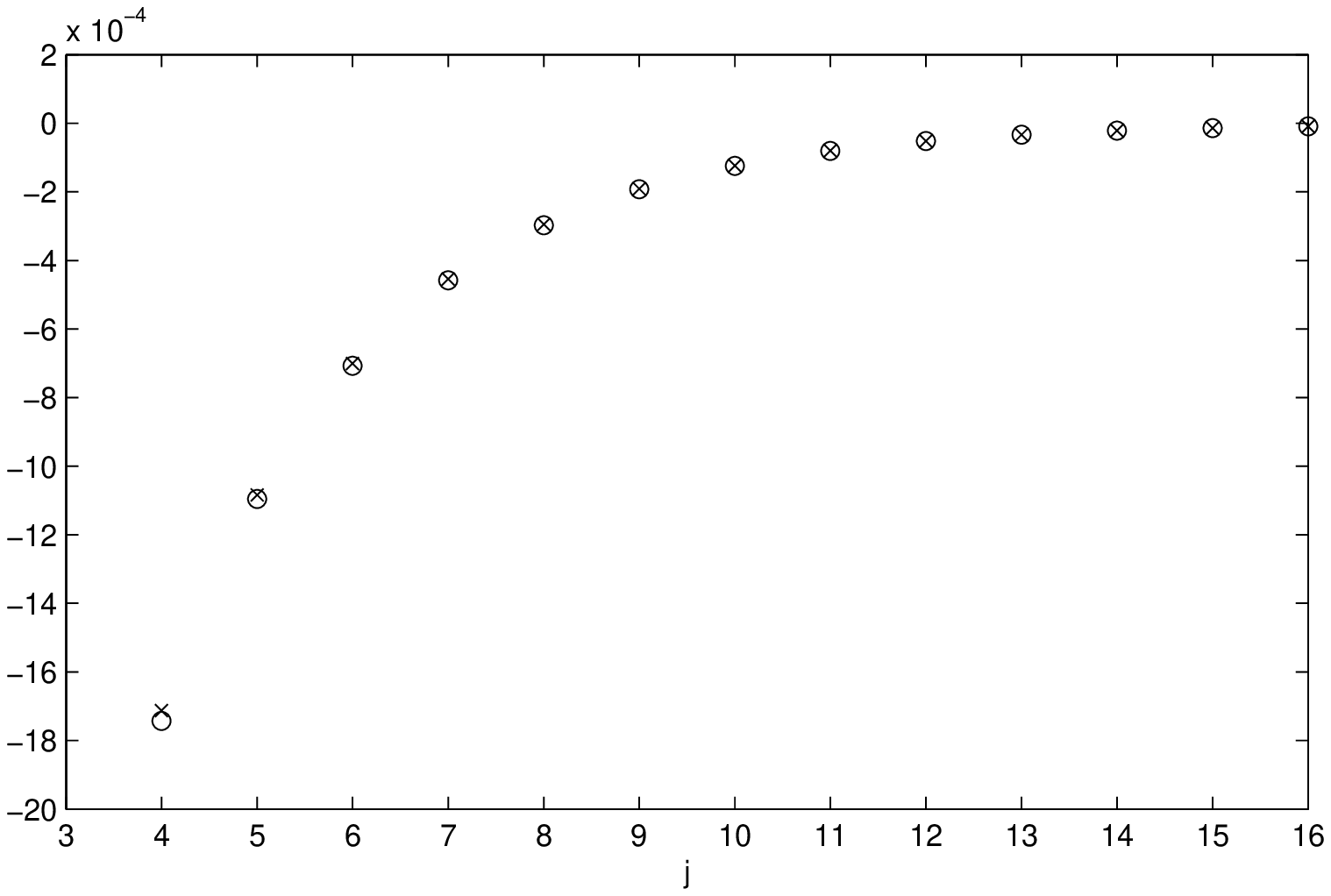,width=0.47\textwidth}}
\subfigure[]{\epsfig{figure=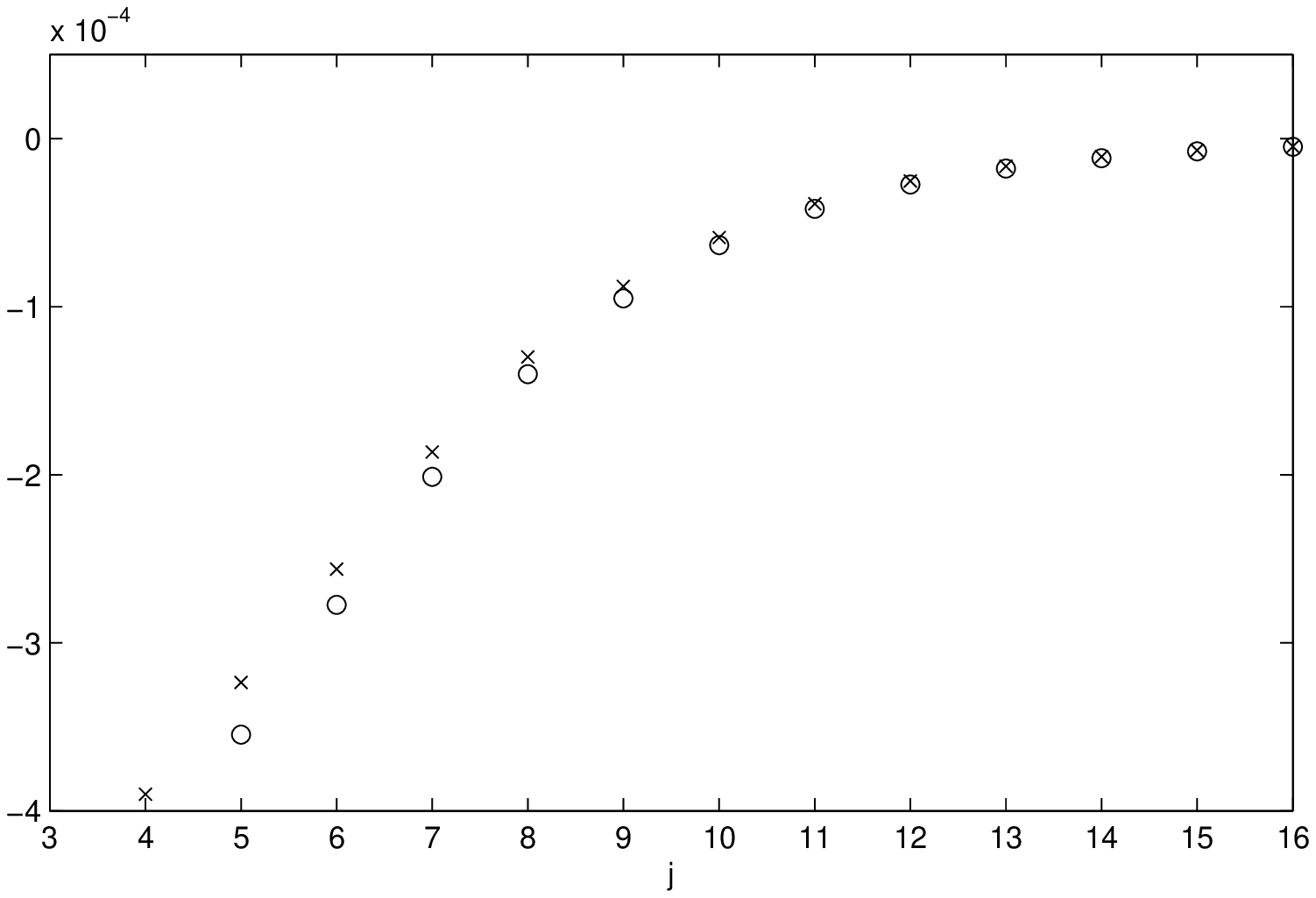,width=0.47\textwidth}}
\caption{The left figure shows the values of $f_i'(0)$ ($\times$'s) and $f_i(T_i)/T_i$ (open circles), and the right figure shows the values of $g_i'(0)$ ($\times$'s) and $g_i(T_i)/T_i$ (open circles) on each pass the trajectory makes past an equilibrium ($i$) for the integration shown in figure~\ref{fig:int1}.}
\label{fig:fip}
\end{center}
\end{figure}

The third assumption is that that $f_i(T_i)\approx T_if_i'(0)$, and $g_i(T_i)\approx T_ig_i'(0)$. Note that in the recurrence relations~\eqref{eq:rr}, the terms in $f_i'(0)$ only appear in the combination $\lambda+f_i'(0)$ (similarly with $g_i'(0)$ in the combination $-\mu+g_i'(0)$). Therefore, we only need to show that the difference between $f_i'(0)$ and $f_i(T_i)/T_i$ is much smaller than $\lambda$ to justify our assumption (and similar for the $g_i$). 
For the integration we perform for figure~\ref{fig:int1} we compute the values of $f_i'(0)$, $T_i$ and $f_i(T_i)$ (and the corresponding values for $g_i$) on each pass the trajectory makes past an equilibrium. We plot these values in figure~\ref{fig:fip}. It can be seen that the difference between $f_i'(0)/T_i$ and $f_i(T_i)$ (and between $g_i'(0)/T_i$ and $g_i(T_i)$)  is clearly very small, and is much less than $\lambda=0.7$ ($\mu=0.7012$) for this example. 

\section{Discussion}
\label{sec:conc}

We have shown that a time-delayed feedback control mechanism similar to that first introduced by Pyragas can be used to stabilise periodic orbits of arbitrarily large period, specifically those resulting from a resonant bifurcation from a heteroclinic cycle. Our analytical results are based on a analysis of the stabilisation of orbits near the Guckenheimer--Holmes cycle. These results are asymptotic, that is, they are correct in the limit of the periodic orbit being close to the heteroclinic cycle. However, in comparison with numerical results (which conversely, are much harder to obtain when the orbit is close to the cycle due to the long period of the orbit), the results actually agree for some large(ish) range of parameters away from the bifurcation point.

It should also be possible to extend this analysis so that it applies to resonant bifurcations from higher dimensional heteroclinic cycles. However, care may need to be taken with the transverse eigenvalues.

As the resonant bifurcation is approached, the period of the bifurcating periodic orbit grows like $1/\nu$, where $\nu$ is the bifurcation parameter. This is in contrast to the homoclinic bifurcation, in which case the bifurcating periodic orbit has a period which grows like $-\log\nu$. This difference in scaling between the growth rate of the periods of the orbits indicates that the results of adding similar time-delayed feedback near a subcritical homoclinic bifurcation may be quite different to the results given here. Work on this problem is ongoing.

\section*{Acknowledgments}

The author would like to thank Mary Silber for many useful discussions regarding this work, and David Barton for assistance using \textsc{dde-biftool}. Two anonymous referees also provided some helpful comments. This work was supported in part by grant NSF-DMS-0709232.


\begin{thebibliography}{10}

\bibitem{Pyr92} K.~Pyragas, Continuous control of chaos by
 self-controlling feedback, \emph{Phys. Letts. A},  {\bf 170} (1992),
 421--428.

\bibitem{PT93} K.~Pyragas and A.~Tama\v{s}evi\v{c}ius, Experimental
  control of chaos by delayed self-controlling feedback,
  \emph{Phys. Letts. A},  {\bf 180} (1993), 99.

\bibitem{GSCS94} D.~J.~ Gauthier, D.~W.~ Sukow, H.~M.~Concannon and
J.\ E.\ S.\ Socolar, Stabilizing unstable periodic orbits in a fast diode
resonator using continuous time-delay autosynchronization,
\emph{Phys. Rev. E}, {\bf 50} (1994), 2343.

\bibitem{BDG94} S.\ Bielawski, D.\ Derozier and P.\ Glorieux, Controlling
unstable periodic orbits by a delayed continuous feedback,
\emph{Phys. Rev. E},  {\bf 49} (1994), R971.

\bibitem{PBA96} Th.\ Pierre, G.\ Bonhomme and A.\ Atipo  
Controlling the Chaotic Regime of Nonlinear Ionization Waves using
the Time-Delay Autosynchronization Method,
\emph{Phys. Rev. Lett.},  {\bf 76} (1996), 2290.

\bibitem{FSK02} T.~Fukuyama, H.~Shirahama and Y.~Kawai,
Dynamical control of the chaotic state of the current-driven
ion acoustic instability in a laboratory plasma using delayed feedback,
\emph{Physics of Plasmas}, {\bf 9} (2002),  4525.

\bibitem{SBFHLM93} F.~W.~Schneider, R.~Blittersdorf, A.~F\"{o}rster,
T.~Hauck, D.~Lebender and J.~M\"{u}ller, 
Continuous Control of Chemical Chaos by Time Delayed Feedback,
\emph{J. Phys. Chem.}, {\bf 97} (1993),  12244.

\bibitem{LFS95} A.~Lekebusch, A.~F\"{o}rster and F.~W.~Schneider, 
Chaos Control in an Enzymatic Reaction, \emph{J. Phys. Chem.}, 
{\bf 99} (1995), 681.

\bibitem{BS96b} M.~E.~Bleich, J.~E.~S.~Socolar, Controlling
  spatiotemporal dynamics with time-delay feedback \emph{Phys. Rev. E},
   {\bf 54(1)} (1996).

\bibitem{MS04} K.~Montgomery and M.~Silber, Feedback Control of
  Traveling Wave Solutions of the Complex Ginzburg Landau Equation,
  \emph{Nonlinearity},  {\bf 17(6)} (2004), 2225-2248.

\bibitem{LYH96} W.~Lu, D.~Yu, R.~G.~Harrison, Control of patterns in
  spatiotemporal chaos in optics, \emph{Phys. Rev. Letts.}, {\bf
  76(18)} (1996), 3316--3319.

\bibitem{PS06} C.~M.~Postlethwaite and M.~Silber, Spatial and temporal 
feedback control of traveling wave solutions of the two-dimensional 
complex Ginzburg--Landau equation, \emph{Physica D},  {\bf 236} (2007), 65--74.

\bibitem{Pyr06} K.~Pyragas, Delayed feedback control of chaos,
 \emph{Phil. Trans. R. Soc. A},  {\bf 364} (2006), 2309--2334.

\bibitem{HS98} J.~Hofbauer and K.~Sigmund, \emph{Evolutionary Games and
  Population Dynamics} (1998), CUP.

\bibitem{Kru97} M.~Krupa, Robust heteroclinic
cycles {\it J. Nonlinear Sci.},  {\bf 7} (1997), 129-176.

\bibitem{KS94} V.~Kirk and M.~Silber, A Competition between
heteroclinic cycles {\it Nonlinearity},  {\bf 7} (1994), 1605-1621.

\bibitem{KM04} M.~Krupa and I.~Melbourne, Asymptotic stability of heteroclinic cycles in systems with symmetry. II, {\it Proc. Roy. Soc. Ed. A},  {\bf 134} (2004), 1177--1197 

\bibitem{Chos97} P.~Chossat, M.~Krupa, I.~Melbourne and
A.~Scheel, Transverse bifurcations of homoclinic cycles {\it Physica
D},  {\bf 100} (1997), 85-100.

\bibitem{CDF90} S.-N.~Chow, B.~Deng and B.~Fielder, Homoclinic bifurcation at resonant eigenvalues, \emph{J. Dyn.
Diff. Eq.},  \textbf{2} (1990), 177-244.

\bibitem{PD06} C.~M.~Postlethwaite and J.~H.~P.~Dawes, A codimension-two resonant bifurcation from a heteroclinic cycle with complex eigenvalues. {\it Dynamical Systems: An International Journal},  21(3) (2006), 313-336.

\bibitem{GH88} J.~Guckenheimer and P.~Holmes, Structurally
stable heteroclinic cycles. {\it Math. Proc. Camb. Phil. Soc.}, 
{\bf 103} (1988), 189-192.

\bibitem{KM95} M.~Krupa and I.~Melbourne, Asymptotic stability of heteroclinic cycles in systems with symmetry, \emph{Erg. Th. Dyn. Sys.}, {\bf 15} (1995) 121--147.

\bibitem{Field96} M.~J.~Field {\it Lectures on bifurcations, dynamics and
  symmetry}, Pitman Research Notes in Mathematics, {\bf 356} (1996).

\bibitem{PS07b} C.~M.~Postlethwaite and M.~Silber, Stabilizing unstable
 periodic orbits in the Lorenz equations using time-delayed feedback control,
 \emph{PRE},  {\bf 76} (2007), 056214 

\bibitem{Fie06} B.~Fiedler, V.~Flunkert, M.~Georgi, P.~Hovel and
  E.~Scholl, Refuting the odd number limitation of time-delayed
  feedback. \emph{Phys. Rev. Lett.},  {\bf 98} (2007), 114101.

\bibitem{ddebiftool} K.~Engelborghs, T.~Luzyanina, G.~Samaey, \textsc{DDE-BIFTOOL} v. 2.00: a Matlab package for bifurcation analysis of delay differential equations, Technical Report TW-330, Department of Computer Science, K.~U.~Leuven, Leuven, Belgium, 2001.


\end{thebibliography}
\end{document}